\documentclass[twocolumn,11pt]{article}

\usepackage{amsmath}
\usepackage{graphicx,url}

\advance\oddsidemargin by -1in  
\advance\evensidemargin by -1in 

\advance\topmargin by -1.5in      

\begin{document}

\title{A novel changepoint detection algorithm}

\author{Allen B. Downey}

\date{}

\maketitle

\begin{abstract}

\textbf{Note added in version 2 (2026): this paper has been superseded.}
The probability it computes, that each index is the most recent
changepoint, is the run-length posterior of Bayesian online
changepoint detection (R.~P. Adams and D.~J.~C. MacKay,
arXiv:0710.3742, 2007), which computes it exactly and efficiently;
see also P.~Fearnhead, \emph{Statistics and Computing} 16:203--213,
2006.  The implementation used for the results in this paper
contained errors, so those results should not be relied on.  A
corrected implementation, tests, and a description of the errors are
at \url{https://github.com/AllenDowney/ChangePoint}.  The rest of the
paper is unchanged from version 1.

\medskip

We propose an algorithm for simultaneously detecting and locating
changepoints in a time series, and a framework for predicting the
distribution of the next point in the series.  The kernel of the
algorithm is a system of equations that computes, for each index $i$,
the probability that the last (most recent) change point occurred at
$i$.  We evaluate this algorithm by applying it to the change point
detection problem and comparing it to the generalized likelihood ratio
(GLR) algorithm.  We find that our algorithm is as good as GLR, or
better, over a wide range of scenarios, and that the advantage
increases as the signal-to-noise ratio decreases.

\end{abstract}

\section{Introduction}
\label {intro}

Predicting network performance characteristics is useful for a variety
of applications.  At the protocol level, predictions are used to
set protocol parameters like timeouts and send rates
\cite{noureddine02transmission}.  In middleware, they are used for
resource selection and scheduling \cite{wolski99network}. And at the
application level, they are used for predicting transfer times and
choosing among mirror sites.

Usually these predictions are based on recent measurements, so their
accuracy depends on the constancy of network performance.  For some
network characteristics, and some time scales, Internet measurements
can be described with stationary models, and we expect simple
prediction methods to work well.  But many network measurements are
characterized by periods of stationarity punctuated by abrupt
transitions \cite{zhang01constancy} \cite{karagiannis04nonstationary}.
On short time scales, these transitions are caused by variations in
traffic (for example, the end of a high-throughput transfer).  On
longer time scales, they are caused by routing changes and
hardware changes (for example, routers and links going down and coming
back up).

These observations lead us to explore statistical techniques for
changepoint detection.  Changepoint analysis is based on the
assumption that the observed process is stationary during certain
intervals, but that there are changepoints where the parameters of the
process change abruptly.  If we can detect changepoints, we know
when to discard old data and estimate the new parameters.

Thus, changepoint analysis can improve the accuracy of predictions.
It can also provide meta-predictions; that is, it can be used to
indicate when a prediction is likely to be accurate.

Statisticians have done extensive work on the general problem of
changepoint detection.  Variations on the changepoint problem
\cite{bhattacharya94some} include:

\begin{description}

\item[Changepoint detection:] The simplest version of the changepoint
problem is testing the hypothesis that a changepoint has
occurred.  Elementary algorithms for change detection include control
charts, CUSUM algorithms and likelihood ratio tests
\cite{basseville93detection}.

\item[Location estimation:] For some applications it is important
to estimate the location (time) of the changepoint.  More generally, it
might be useful to find a confidence set or an interval of times
that contain a changepoint at some level of confidence.

\item[Tracking:] The goal of the tracking problem is to partition a
time series into stationary intervals and estimate the parameters of
the process in each interval.  A simple approach is to use hypothesis
testing to detect a changepoint, estimate the location of the change(s),
and then use conventional techniques to estimate the parameters of the
process in each interval.

\end{description}

For each of these problems, there is an alternative Bayesian
formulation.  Instead of testing (and possibly rejecting) the
hypothesis that a changepoint has occurred, the Bayesian approach is
to compute the subjective probability of the hypothesis.  Instead of
estimating the location, the Bayesian approach is to compute $P(i)$,
the probability that there is a changepoint at index $i$.  The
Bayesian version of the tracking problem is to compute the probability
of a changepoint at a given time and the conditional distribution of
the parameters (conditioned on the location).

In this paper, we take the Bayesian approach and apply it to the
tracking problem.  Our algorithm can also be used for changepoint
detection and location estimation.

\subsection{Online or off?}

Changepoint detection algorithms are sometimes classified as
``online'' or ``offline'', but this is not an intrinsic characteristic
of the algorithm; it is a description of the deployment.  Online
algorithms run concurrently with the process they are monitoring,
processing each data point as it becomes available, with the real-time
constraint that processing should complete before the next data point
arrives.  Offline algorithms consider the entire data set at once, and
there is no real-time constraint on the run time.

So in most cases, ``online'' means ``fast enough to keep up''.
Sometimes it means ``incremental'', in the sense that when a new
data point arrives, the algorithm can perform an update, based on
previously stored results, faster than it could recompute from
scratch.  Incremental algorithms are particularly appealing in
embedded applications, where processing power is at a premium, but it
is not generally necessary for online algorithms to be incremental.

``Online'' and ``offline'' are probably best used to describe the
problem formulation rather than the algorithm.  In the online
scenario, the goal is to detect a changepoint as soon as possible
after it occurs, while minimizing the rate of false alarms.  In the
offline scenario, the goal is often to identify all changepoints
in a sequence, and the the criteria are sensitivity (detecting
true changepoints) and specificity (avoiding false positives).

The algorithm we are proposing can be used for both online and offline
problems.  It is incremental, so it lends itself to online deployment;
on the other hand, it requires computation time proportional to $n^2$
for each data point, where $n$ is the number of points in the
current window, so for some applications it may not be fast enough
to satisfy real time requirements.

Nevertheless, since the applications we are considering (making real
time predictions) are online, we evaluate our algorithm by the
criteria of online problems.

\subsection{Prediction with changepoints}

The goal of this work is to generate a predictive distribution from a
time series with changepoints.  For example, imagine that we have
observed $n$ values $x_1 \ldots x_n$, and we want to predict
the value of $x_{n+1}$.  In the absence of changepoints, we might model
the series as a random variable $X$ with a stationary distribution
(and, possibly, autocorrelation structure).  We could use the observed
values to estimate the parameters of $X$ and then use those parameters
to compute the distribution of the next value.

If we know that there is a changepoint, and we assume that there is
no relation between the parameters of the process before and after
the change, then the data before the changepoint are irrelevant
for making predictions.

But in the context of network measurements, we usually don't know when
a changepoint occurs.  We can only infer it from the time series
itself, and we usually don't know with certainty whether it occurred or
exactly when.

There are two ways to handle this uncertainty.  The most common
approach is to compute from the data a functional that is likely to
increase after a changepoint and to compare the resulting values to a
previously-chosen threshold.  If the value exceeds the threshold, we
assume that there was a changepoint and discard older data.  Otherwise
we assume that there was no changepoint.

The alternative, which we explore here, is to use changepoint
probabilities to generate predictive distributions.  For example, if
we think that there is $p$ chance of a changepoint at $i$, then we
can generate two distributions, one that assumes that there was a
changepoint, and one that assumes that there wasn't, and then mix
the distributions using $p$ as a weight.

Therefore, our intermediate goal is to compute
$P(i^+)$, which we define as the probability that $i$ is the index
of the last (most recent) changepoint.  The $+$ notation is intended
to distinguish from $P(i)$, which is the probability that $i$
is the index of a changepoint (but not necessarily the last).

\subsection{Related Work}

In a seminal paper on the tracking problem, Chernoff and Zacks present
a Bayesian estimator for the current mean of a process with abrupt changes
\cite{chernoff64estimating}.
Like us, they start with an estimator that assumes there is at most
one change, and then use it to generate an approximate estimator in
the general case.  Their algorithm makes the additional assumption
that the change size is distributed normally; our algorithm does not
require this assumption.  Also, our algorithm generates a predictive
distribution for the next value in the series, rather than an estimate
of the current mean.

Since then, there has been extensive work on a variety of related
problems, using both Bayesian and non-Bayesian approaches.  More
recent works include extensions to deal with multiple changepoints,
autoregressive processes, and long-range dependent processes.

The algorithm we propose can be extended to detect changes in the
variance as well as the mean of a process.  This kind of changepoint
has received relatively little attention; one exception is recent
work by Jandhyala, Fotopoulos and Hawkins \cite{jandhyala02detection}.

Most recently, Vellekoop and Clark propose a nonlinear filtering
approach to the changepoint detection problem (but not estimation
or tracking) \cite{vellekoop06nonlinear}.

We are aware of a few examples where these techniques have been
applied to network measurements.  Bla\v{z}ek et al explore the use of
change-point algorithms to detect denial of service attacks
\cite{blazek01novel}.  Similarly Deshpande, Thottan and Sikdar use
non-parametric CUSUM to detect BGP instabilities
\cite{deshpande04early}.

In the context of large databases, Kifer, Ben-David and Gehrke propose
an algorithm for detecting changes in data streams
\cite{kifer04detecting}.  It is based on a two-window paradigm, in
which the distribution of values in the current window is compared to
the distribution of values in a past reference window.  This approach
is appropriate when the number of points between changepoints is
large and alarm delay is not a critical metric.

\subsection{Outline}

Before presenting our algorithm we provide a summary of the techniques
and results we will use from Bayesian statistics
(Section~\ref{bayes}).  Then we develop our algorithm, starting with a
direct method for the case where we know there is exactly one
changepoint, and then using that result to develop an iterative method
for the case where the number of changepoints is unknown
(Section~\ref{estimation}).  That section includes synthetic examples
to demonstrate changes in both mean and variance, and one real series,
the ubiquitous Nile dataset.  Finally, we apply our algorithm to the
changepoint detection problem and compare it with GLR, a standard
algorithm for this problem (Section~\ref{evaluation}).

\section{Bayesian Statistics}
\label{bayes}

This section presents an introduction to Bayesian statistics, and
develops several examples we will use in later sections.

The fundamental idea behind all Bayesian statistics is the
diachronic interpretation of Bayes' Theorem:

\begin{equation}
P(H|E) = \frac{P(E|H) P(H)}{P(E)}
\label{bayeseq}
\end{equation}
where $H$ is a hypothesis and $E$ is a relevant body of evidence.
$P(H|E)$ is the probability of the hypothesis given the evidence,
which is called the {\em posterior probability}.  $P(H)$ is the
probability of the hypothesis before we take account of the
evidence, known as the {\em prior probability}.  $P(E|H)$ is the
probability of seeing the evidence, given that the hypothesis
is true; this term is sometimes called the {\em likelihood}.
Finally, $P(E)$ is the probability of seeing the evidence under
any possible hypothesis, called the {\em normalizing constant}.

If the hypotheses we are interested in are stochastic
processes, it is usually straightforward to compute $P(E|H)$.  The
prior probability $P(H)$ is either based on a previous computation or
reflects a subjective degree of belief in the hypothesis before seeing
the evidence.  The normalizing constant is hard to formulate in general,
but if we can identify a mutually exclusive and complete set of
hypotheses, S, then

\begin{equation}
P(E) = \sum_{H_i \in S} P(E|H_i) P(H_i)
\end{equation}
In practice it is sufficient to compute any {\em likelihood function},
$L(E|H)$, such that

\begin{equation}
L(E|H) = \kappa P(E|H) P(H)
\end{equation}
where $\kappa$ is an arbitrary constant that drops out when we
compute
\begin{equation}
P(H|E) = \frac{L(E|H)}{\sum_{H_i \in S} L(E|H_i)}
\end{equation}

As an example, suppose that $H$ is the hypothesis that a random
variable $X$ is distributed $N(\mu,\sigma^2)$ with known $\mu$ and
$\sigma^2$.  
In a continuous distribution, the probability of a given selection
$X=x$ is not well defined, but since the probability density function
$pdf_X(x)$ is proportional to $P(X=x|H)$, we can use it as a
likelihood function:
\begin{multline}
L(X=x|H) = pdf_X(x) =\\
         \frac{1}{\sigma \sqrt{2 \pi}} 
\exp \left[ -\frac{1}{2} \left( \frac{x - \mu}{\sigma} \right) ^2 \right]
\end{multline}
For a sample $E = x_i$ with $i = 1 \ldots n$
\begin{equation}
L(E|H) = \prod_i pdf_X(x_i)
\end{equation}

\subsection{Posterior distributions}

We have been using the notation $P(A)$ to mean the probability of an
event $A$.  Following the notation of Gelman et al
\cite{gelman03bayesian}, we use $p(x)$ as a shorthand for the
probability density function $pdf_X(x)$.  This notation makes it
convenient to write Bayes' Theorem for distributions.  For example,
if we observe a set of data, $y$, we might want to compute the
distribution of $\theta$, which is the set of parameters of the
process that produced $y$.  Generalizing Bayes' Theorem yields
\begin{equation}
p(\theta|y) = \frac{p(y|\theta) p(\theta)}{p(y)}
\end{equation}
where $p(\theta)$ is the prior distribution
of $\theta$, and $p(\theta|y)$ is the posterior distribution.

As an example, suppose that $y$ is a set of interarrival times between
passengers at a bus stop, and the arrival process is Poisson; in this
case $\theta$ is the single parameter of the Poisson model (the mean
arrival rate), and $p(\theta|y)$ is the distribution of $\theta$,
taking the observed data into account.  In frequentist statistics,
$\theta$ is not a random variable, so it is not meaningful to talk
about its distribution, but in the Bayesian interpretation,
probability can reflect a degree of belief about a non-random
quantity.

As another example, which we will use in Section~\ref{estimation},
suppose that we have a set of observations $y$ that come from a normal
distribution with unknown mean $\mu$ and variance $\sigma^2$.  We
would like to compute posterior distributions for $\mu$ and $\sigma$.
For this problem it is common to use a noninformative prior
distribution that is uniform on ($\mu$, $\log (\sigma)$), in which
case the posterior distribution of $\sigma^2$ is a scaled inverse
chi-square distribution:
\begin{equation}
\label{sigmapost}
\sigma^2|y \sim \mbox{Inv-} \chi ^2(n-1, s^2)
\end{equation}
where $n$ is the number of observations in $y$ and $s^2$ is
the estimated variance,
\begin{equation}
s^2 = \frac{1}{n-1} \sum_{i=1}^n (y_i - \bar{y})^2
\end{equation}
where $\bar{y}$ is the estimated mean $\frac{1}{n} \sum_{i=1}^n y_i$.

Then for a given value of $\sigma$ the conditional posterior
distribution of $\mu$ is Gaussian:
\begin{equation}
\label{mupost}
\mu|y,\sigma^2 \sim N(\bar{y}, \sigma^2/n )
\end{equation}
For the derivation of this result, see Gelman et al
\cite{gelman03bayesian}, pages 74--76.

\section{Estimating $P(i^+)$}
\label{estimation}

Imagine that we have observed a series of $n$ values.  We would like
to estimate $P(i^+)$, the probability that $i$ is the index of the last
(most recent) changepoint, for all $i$.

We start with a simplified version of the problem and work
up to the most general version.
To get started, we assume 

\begin{enumerate}

\item Exactly one changepoint occurred
during the observed interval, and it is equally likely to be
anywhere in the interval.

\item Between changepoints, the distribution of values is normal
with constant mean and variance.

\item Before the changepoint, the mean is known to be $\mu_0$.  After
the changepoint the mean is unknown.

\item The variance is known to be $\sigma^2$ both before and after
the changepoint.

\end{enumerate}

The first assumption is useful because it means that the series of
hypotheses, $i^+$, is mutually exclusive and complete, which makes it
practical to compute the normalizing constant in
Equation~\ref{bayeseq}.  The other assumptions make it easy to compute
a likelihood function.

In the following sections, we present an algorithm for this
simplified problem and then gradually relax the assumptions.

\subsection{Exactly one changepoint}
\label{likefun}

We would like to compute $P(E|i^+)$, which is the probability of
seeing the observed data $E$ given the hypothesis $i^+$.
If we know there is only one changepoint in an interval,
$i^+$ is equivalent to the hypothesis that the only
changepoint is at $i$, so the likelihood function is
\begin{multline}
P(E|i^+, \mu_0, \sigma^2) = \\
\prod_{j=1}^{i} P(x_j | \mu_0, \sigma^2) 
\prod_{j=i+1}^{n} P(x_j | \mu_1, \sigma^2)
\label{likelihood}
\end{multline} 
where $\mu_1$ is the mean after the changepoint.  Since $\mu_1$
is unknown, it has to be estimated from the data $x_{i+1} \ldots x_n$.  Given
the sample mean and variance, we can compute the posterior
distribution $p(\mu_1)$ using Equation~\ref{mupost}, and then
\begin{equation}
P(E|i^+) = \int P(E|i^+, \mu_0, \mu_1, \sigma^2) p(\mu_1) d \mu_1
\end{equation}
It is not practical to evaluate this integral, but since the algorithm
we are proposing is iterative, we can approximate it by
making a random selection from $p(\mu_1)$
during each iteration.

If the changepoint is equally likely to occur anywhere, the prior
probabilities are all equal, so they drop out of the computation.
The set of hypotheses is exclusive and complete, so the normalizing
constant is
\begin{equation}
P(E) = \sum_{i=1}^n P(E|i^+)
\end{equation}

\subsection{Zero or one changepoints}
\label{likefun0}

If we know that there are zero or one changepoints
in the interval, we can generalize this technique by adding
$H_0$, which is the hypothesis that there are no changepoints.

If we assume that the probability of a changepoint at any time is $f$,
then the prior probabilities\footnote{This is an improper
prior that is proportional to the actual probabilities.} are
\begin{equation}
\begin{gathered}
P(i^+) = f (1-f)^{n-1} \\
P(H_0) = (1-f)^n
\end{gathered}
\end{equation}
The likelihood of $H_0$ is
\begin{equation}
P(E|H_0,\mu_0, \sigma^2) = \prod_{j=1}^{n} P(x_j | \mu_0, \sigma^2) 
\end{equation}
Again, we have a set of hypotheses that is 
exclusive and complete, so the normalizing constant is
\begin{equation}
P(E) = P(E|H_0) P(H_0) + \sum_{i=1}^n P(E|i^+) P(i^+)
\end{equation}

\subsection{Iterative algorithm}

As the next step, we drop the assumption that we know that exactly one
changepoint has occurred.  In that case we can't evaluate the $P(i^+)$
directly, but we can write the following system of equations for the
$P(i^+)$ and the $P(i^{++})$, where $i^{++}$ is the hypothesis
that the {\em second-to-last} changepoint is at $i$.  This system is
the kernel of the algorithm we are proposing.
\renewcommand{\P}{\ P}
\newcommand{\jc}{H_\oslash}
\begin{equation}
\begin{split}
&P(i^+) = P(i^+ | \jc) \P(\jc) + \sum_{j<i} P(i^+ | j^{++}) \P(j^{++})   \\
&P(i^{++}) = \sum_{k>i} P(i^{++} | k^+) \P(k^+) 
\label{cpp}
\end{split}
\end{equation}
where
\begin{itemize}
\item $\jc$ is the hypothesis that there are fewer than two
changepoints during the observed interval (the subscript is meant
to look like ``zero or one''):
\begin{equation}
P(\jc) = 1 - \sum P(j^{++})
\end{equation}
\item $P(i^+ | j^{++})$ is the probability that $i$ is the last
changepoint, given that $j$ is the second-to-last.  Computing this
probability reduces to the problem in Section~\ref{likefun}
because if $j$ is the second-to-last changepoint,
we know that there is exactly one changepoint between $j$ and $n$.
\item $P(i^+ | \jc)$ is the probability that $i$ is the last
changepoint, given that there are fewer than two changepoints in the
interval.  This reduces to the problem in Section~\ref{likefun0},
where we know that there are zero or one changepoints.
\item $P(i^{++} | k^+)$ is the probability that $i$ is the
second-to-last changepoint given that $k$ is the last.  If $k$ is a
changepoint, then all data after time $k$ is irrelevant, so $P(i^{++}
| k^+) = P_k(i^+)$, where $P_k$ indicates the probability that was
computed at time $k$.  If we store previous results, we don't
have to compute this value; we can just look it up.
\end{itemize}
At each time step, we have to recompute $P(i^+ | j^{++})$ and $P(i^+ |
\jc)$.  Then we estimate a solution to the system of equations for
$P(i^+)$ and $P(i^{++})$ by Jacobi's method.
Since the result from the previous time
step is a good starting approximation, one iteration
is usually sufficient.

\subsection{Time and space requirements}

After we have seen and processed $n$ data points, we have accumulated
the following data:

\begin{itemize}

\item $P_k(i^+)$ for all $k \le n$ and $i\le k$, which is roughly
$n^2/2$ entries.

\item The partial sums
\begin{equation}
\sum_{j=i}^{k} x_j \quad \mbox{and} \quad \sum_{j=i}^{k} x_j^2 
\label{partialsum}
\end{equation}
for all $k \le n$ and $i\le k$.  In total we have to store
roughly $n^2$ partial sums.

\end{itemize}

When we see the next data point, $x_{n+1}$, we have to 

\begin{itemize}

\item Update the partial sums (Equation~\ref{partialsum}).

\item Compute the likelihood functions (Equation~\ref{likelihood}).

\item Compute the probabilities $P_{n+1}(i^+)$ and $P_{n+1}(i^{++})$ 
for all $i$ (Equation~\ref{cpp}).

\end{itemize}

Updating the partial sums is linear in $n$, but computing the
likelihoods and probabilities takes time proportional to $n^2$.

This complexity is a drawback of this algorithm in an application
where the time between changepoints is long.  But if there are fewer
than 1000 timesteps between changepoints, this algorithm is feasible
for online applications with modest real time constraints ($<10$
timesteps per second).

\subsection{An example}

\begin{figure}[tb]
\includegraphics[width=3in]{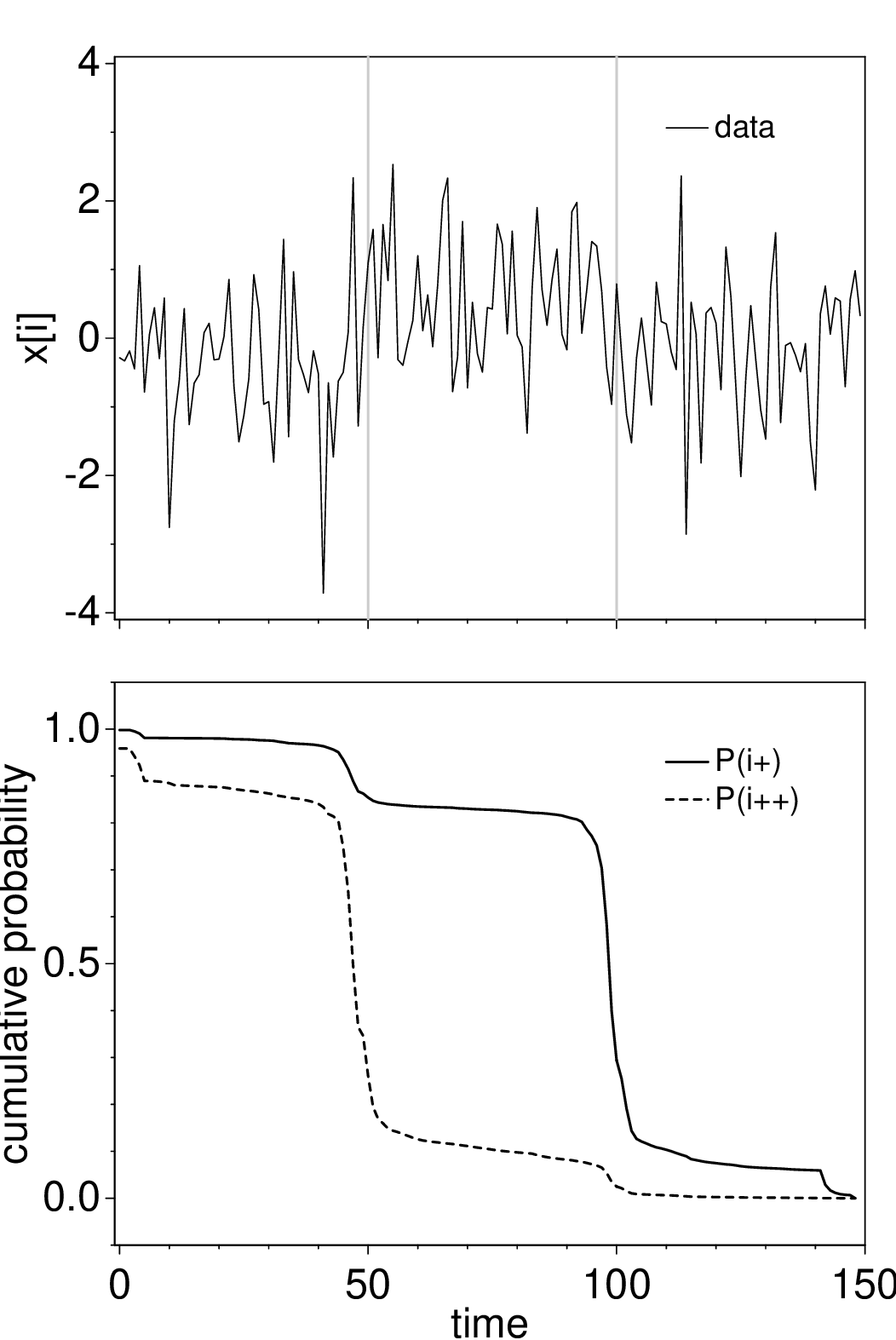}
\caption{A time series with two changepoints (marked with gray lines),
and estimated cumulative probabilities $P(i^+)$ and $P(i^{++})$.}
\label{example}
\end{figure}

Figure~\ref{example} shows a time series with changepoints
at $t=50$ and $t=100$.  At the first changepoint, $\mu$ shifts
from -0.5 to 0.5.  At the second changepoint, it shifts to 0.
Throughout, $\sigma$ is 1.0.

The bottom graph shows the cumulative sums of
$P(i^+)$ and $P(i^{++})$, estimated at $t=150$.
Appropriately, $P(i^+)$ shows a large mode near $t=100$, which
actually is the most recent changepoint.  There is a smaller
mode near $t=50$, the location of the second-most-recent changepoint,
and another near $t=142$, which is due to chance.

The estimated probabilities are not normalized, so they don't
always add up to one.  In this case, the sum of $P(i^{++})$
for all $i$ is $0.96$, which indicates a 4\% chance
that we have not seen two changepoints yet.

\subsection{Generalization to unknown $\mu_0$, $\sigma^2$}

\begin{figure}[tb]
\includegraphics[width=3in]{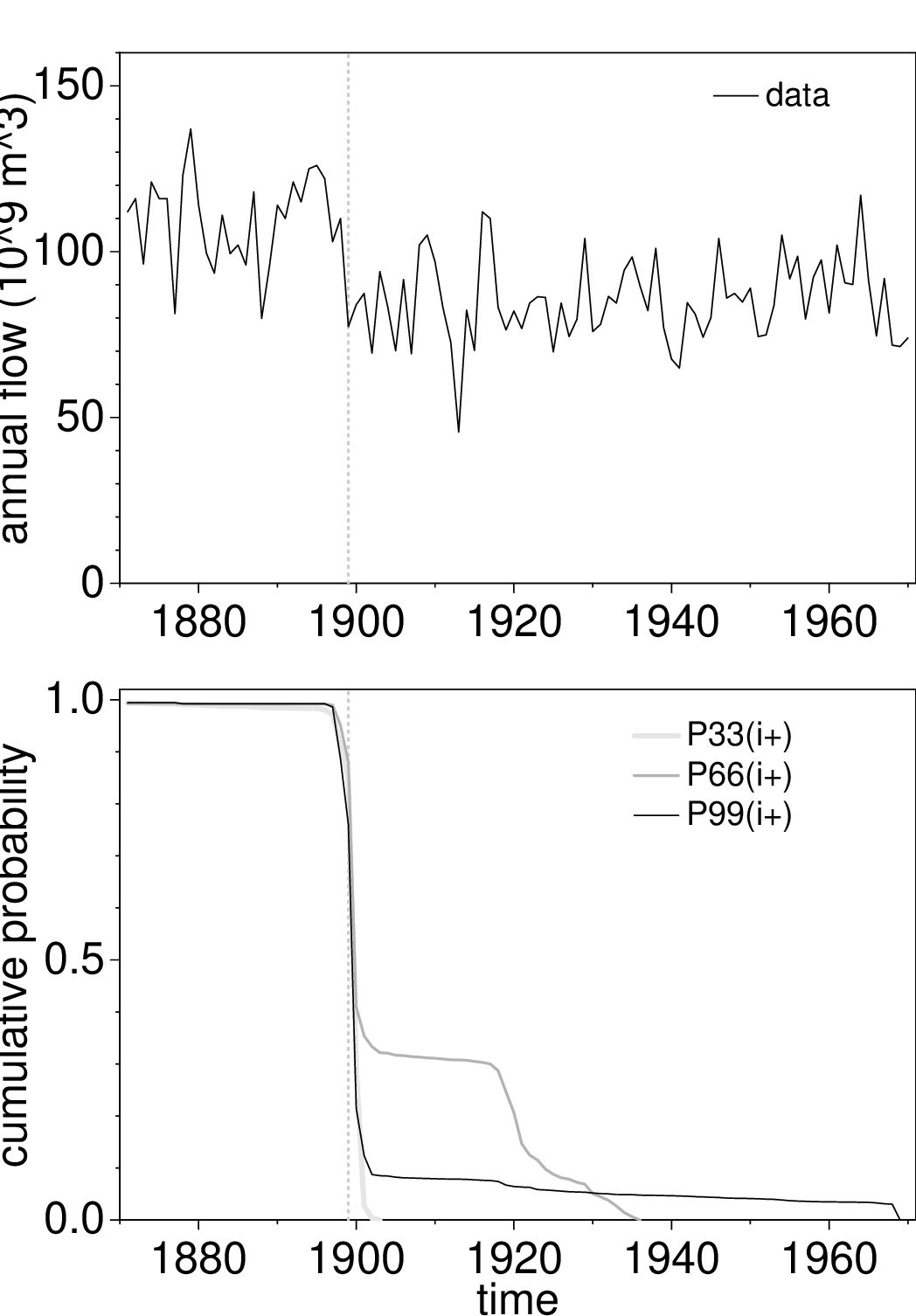}
\caption{Average annual flow in the Nile river at Aswan from 1871 to
1970, and the estimated $P(i^+)$ at $k = 33, 66, 99$.}
\label{nile}
\end{figure}

A feature of the proposed algorithm is that it extends easily
to handle the cases when $\mu_0$ or $\sigma^2$, or both, are
unknown.  There are two possible approaches.

The simpler option is to estimate $\mu_0$ and/or $\sigma^2$ from the
data and then plug the estimates into Equation~\ref{likelihood}.  This
approach works reasonably well, but because estimated parameters are
taken as given, it tends to overestimate the probability of a
changepoint.

An alternative is to use the observed $\bar{y}$ and $s^2$ to compute
the posterior distributions for $\mu_0$ and/or $\sigma^2$ using
Equations~\ref{sigmapost} and ~\ref{mupost}, and then make a random choice
from those distributions (just as we did for $\mu_1$ in
Section~\ref{likefun}).

As an example, we apply our algorithm to the ubiquitous Nile data,
a standard example in changepoint detection since Cobb's seminal
paper in 1978 \cite{cobb78problem}.  This dataset records the 
average annual flow in the Nile river at Aswan from 1871 to 1970.

Figure~\ref{nile} shows this data along with the estimated probability
$P(i^+)$ computed at three different times (after 33, 66 and 99
years).  The salient feature at all three times is a high
probability of a changepoint near 1898, which is consistent with the
conclusion of other authors working with this data.  At $k=66$, there
is also a high probability of a changepoint near 1920, but by $k=99$
that hypothesis has all but vanished.  Similarly, at $k=99$, there is
a small probability that the last four data points, all on the low
side, are evidence of a changepoint.

As time progresses, this algorithm is often quick to identify a
changepoint, but also quick to ``change its mind'' if additional data
fail to support the first impression.  Part of the reason for this
mercurial behavior is that $P(i^+)$ is the probability of being the
{\em last} changepoint; so a recent, uncertain changepoint might have
a higher probability than a more certain changepoint in the past.

\subsection{Detecting changes in variance}
\label{varchange}

\begin{figure}[tb]
\includegraphics[width=3in]{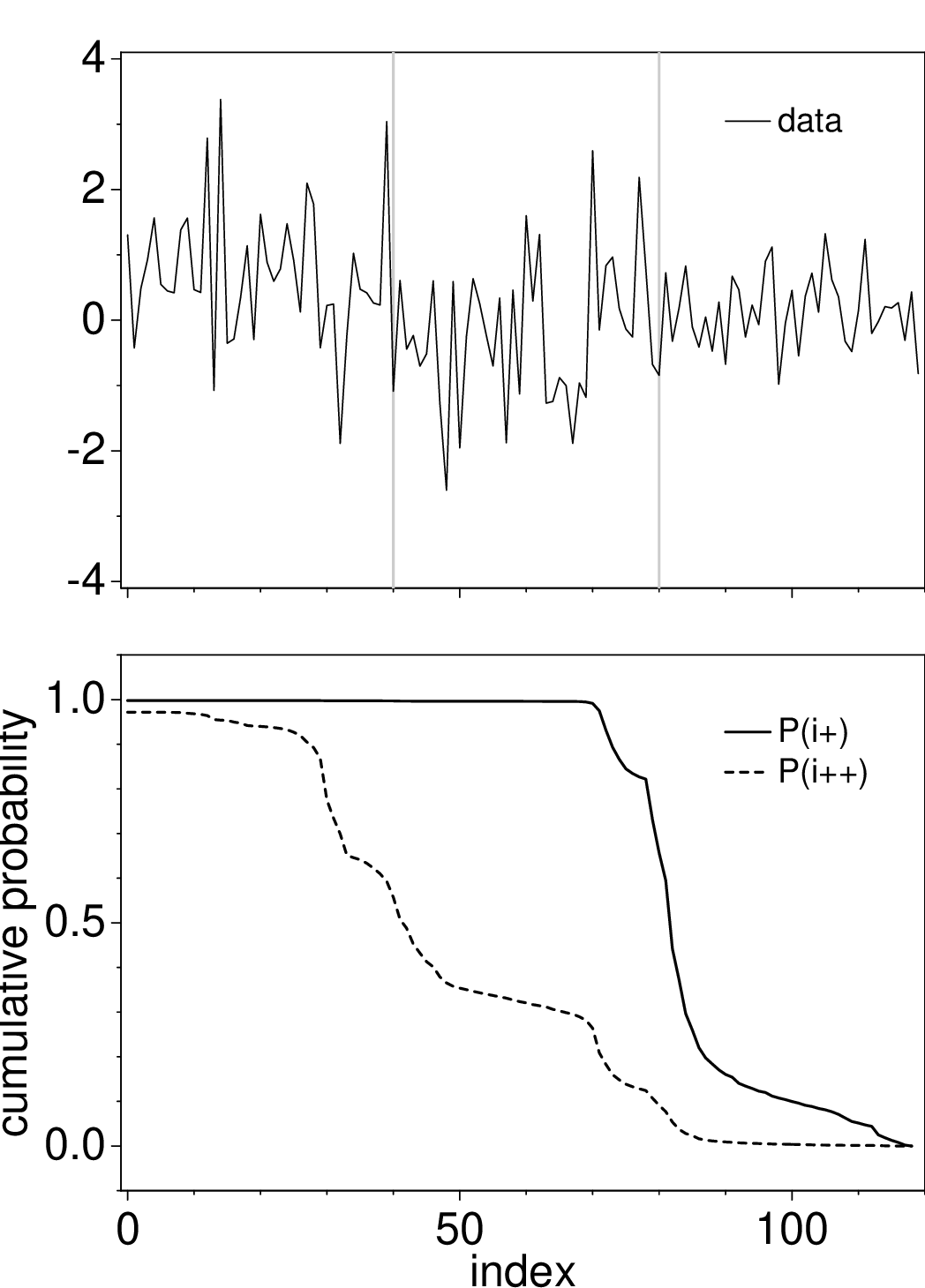}
\caption{A time series with a change in mean and a change
in variance.}
\label{var}
\end{figure}

To detect changes in variance, we can generalize the algorithm further
by estimating $\sigma$ separately before and after the changepoint.
In this case, the likelihood function is
\begin{multline}
P(X|i^+) = \\
\prod_{j=1}^{i} P(x_j | \mu_0, \sigma_0^2) 
\prod_{j=i+1}^{n} P(x_j | \mu_1, \sigma_1^2)
\end{multline}
where $\mu_0$ and $\sigma_0^2$ are drawn from the posterior
distribution for the mean and variance before the changepoint,
and likewise $\mu_1$ and $\sigma_1^2$ after the breakpoint.

Figure~\ref{var} shows a time series with a change in mean
followed by a change in variance.  Before the first changepoint, $\mu,
\sigma = (1, 1)$; after the first changepoint, $\mu, \sigma = (0, 1)$; and
after the second changepoint, $\mu, \sigma = (0, 0.5)$.

Correctly, the estimated $P(i^+)$ indicate a high probability
near the last changepoint.  The estimated $P(i^{++})$ indicate less
certainty about the location of the second-to-last changepoint.

For this example the generalized algorithm works well, but its
behavior is erratic when a series of similar values appear in the
time series.  In that case the estimated values of  $\sigma_1^2$ 
tend to be small, so the computed probability of a changepoint
is unreasonably high.  We are considering ways to mitigate this
effect without impairing the sensitivity of the algorithm
too much.

\section{Evaluation}
\label{evaluation}

The examples in the previous section show that the estimated
values are reasonable, but not that they are accurate or more
useful than results from other algorithms.  Since there are
no previous algorithms for computing $P(i^+)$, there is no
direct basis for comparison, but there are two
evaluations we could perform:

\begin{itemize}

\item There are Bayesian methods that estimate $P(i)$; that is,
the probability that there is a changepoint at any time $i$.
From this we can compute $P(i^+)$ and compare to our algorithm.
We would like to pursue this approach in future work.

\item Changepoint detection algorithms are often used ``online'' to
raise an alarm when the evidence for a changepoint reaches a certain
threshold.  These algorithms are relatively simple to implement, and
the criteria of merit are clear, so we start by evaluating our
algorithm in this context.

\end{itemize}

The goal of online detection is to raise an alarm as quickly
as possible after a changepoint has occurred while minimizing
the frequency of false alarms.

Basseville and Nikiforov \cite{basseville93detection} propose a
criterion for comparing algorithms in the case where an actual
changepoint is equally likely to occur at any time, so the time of the
first changepoint, $t_0$ is distributed geometrically.

During any given run, an algorithm has some probability of
raising an alarm before $t_0$; this false alarm probability is
denoted $\alpha$.  Then, for a constant value of $\alpha$, the
goal is to minimize the {\em mean delay}, $\tau = t_a - t_0 +1$,
where $t_a$ is the time of the alarm.  In our experiements, we
use a trimmed mean, for reasons explained below.

\subsection{GLR}

The gold standard algorithm for online changepoint detection is
CUSUM (cumulative sum), which is optimal, in the sense of minimizing
mean delay, when the parameters of the distribution
are known before and after the changepoint \cite{basseville93detection}.

When the parameters after the changepoint are not known, CUSUM
can be generalized in several ways, including Weighted CUSUM and
GLR (generalize likelihood ratio).

For the experiments in this section, we consider the case where the
distribution of values before the changepoint is normal with known
parameters $\mu_0$, $\sigma^2$; after the changepoint, the
distribution is normal with unknown $\mu_1$ and the same known
variance.  We compare the performance of our algorithm to GLR,
specifically using the decision function $g_k$ in
Equation 2.4.40 from Basseville and
Nikiforov \cite{basseville93detection}.

The decision function $g_k$ is constructed so that its expected value
is zero before the changepoint and increasing afterwards.  When
$g_k$ exceeds the threshold $h$, an alarm is raised, indicating that
a changepoint has occurred (but not when).  The threshold $h$ is
a parameter of the algorithm that can be tuned to trade off between
the probability of a false alarm and the mean delay.

\subsection{CPP}

It is straightforward to adapt our algorithm so that it is comparable
with GLR.  First, we adapt the likelihood function to use the known
values of $\mu_0$ and $\sigma$, so only $\mu_1$ is unknown.  Second, we
use the decision function:

\begin{equation}
g_k = \sum_{i=0}^{k} P(i^+)
\end{equation}

This $g_k$ is the probability that a changepoint has
occurred somewhere before time $k$.  Our algorithm provides additional
information about the location of the changepoint, but this reduction
eliminates that information.

We call this adapted algorithm CPP, for ``change point probability'',
since the result is a probability (at least in the Bayesian sense)
rather than a likelihood ratio.

Both GLR and CPP have one major parameter, $h$, which has a strong
effect on performance, and a minor parameter that has little effect
over a wide range.  For GLR, the minor parameter is the minimum change
size, $\nu_{min}$.  For CPP, it is the prior probability of a
changepoint, $f$.

\subsection{Results}

In our first experiment, $\mu_0, \sigma = (0, 1)$ and $\mu_1 = 1$, so
the signal to noise ratio is $S/N = |\mu_1 - \mu_0| / \sigma = 1$.  In
the next section we will look at the effect of varying $S/N$.

In each trial, we choose the actual value of $t_0$ from a
geometric distribution with parameter $\rho$, so
\begin{equation}
P(t_0 = n) = (1 - \rho)^{n-1} \rho
\end{equation}
We set $\rho = 0.02$, so the average value of $t_0$ is 50.

Each trial continues until an alarm is raised; the alarm time is
denoted $t_a$.  If $t_a \le t_0$, it is considered a false alarm.
Otherwise, we compute the delay time $t_a - t_0 + 1$.

To compute mean delay, we average over the trials that did not raise a
false alarm.  We use a trimmed mean (discarding the largest and
smallest 5\%) for several reasons: (1) the trimmed mean is generally
more robust, since it is insensitive to a small numbers of outliers, (2)
very short delay times are sometimes ``right for the wrong reason'';
that is, the algorithm was about to generate a false
positive when the changepoint occurred, and (3) when $t_0$ is small,
we see very few data points before the changepoint and $t_a$ can be
arbitrarily large.  In our trials, we quit when $k > t_0 + 100$ and
set $t_a = \inf$.  The trimmed mean allows us to generate a
meaningful average, as long as fewer than 5\% of the trials go ``out
of bounds''.

\begin{figure}[tb]
\includegraphics[width=3in]{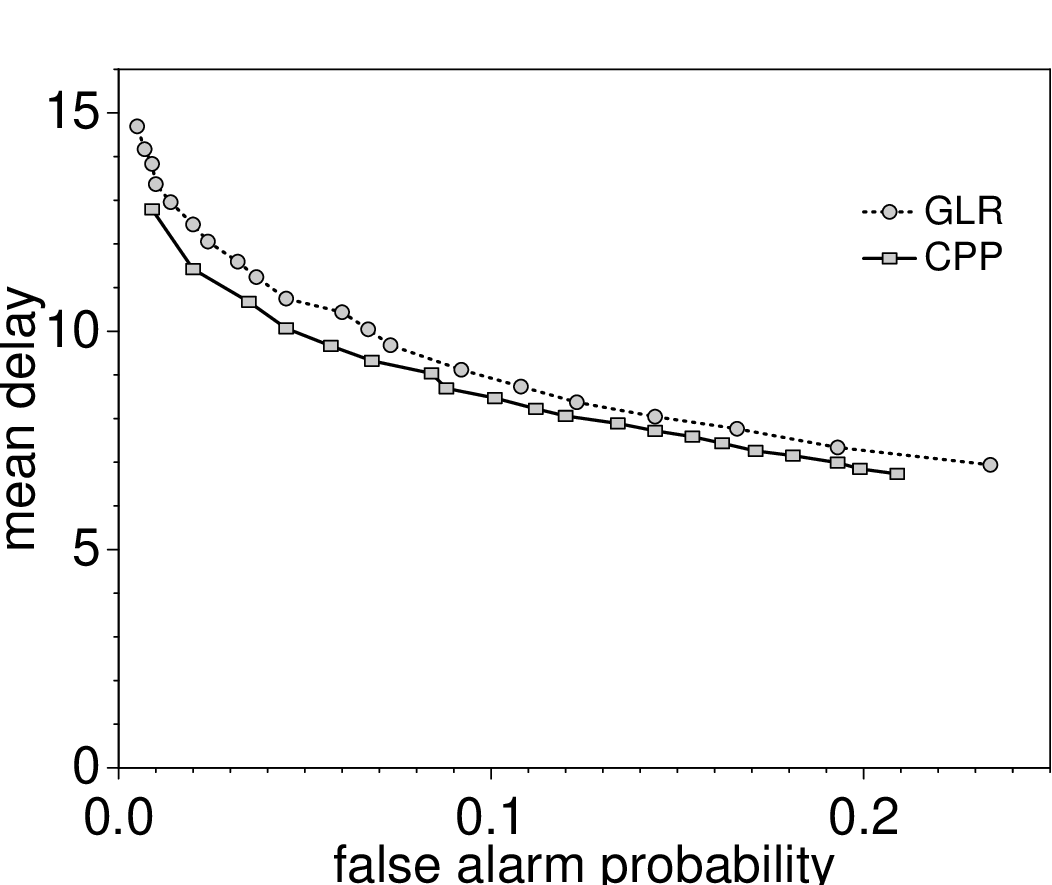}
\caption{Mean delay versus false alarm probability for
a range of threshold values.}
\label{matfar}
\end{figure}

\begin{figure}[tb]
\includegraphics[width=3in]{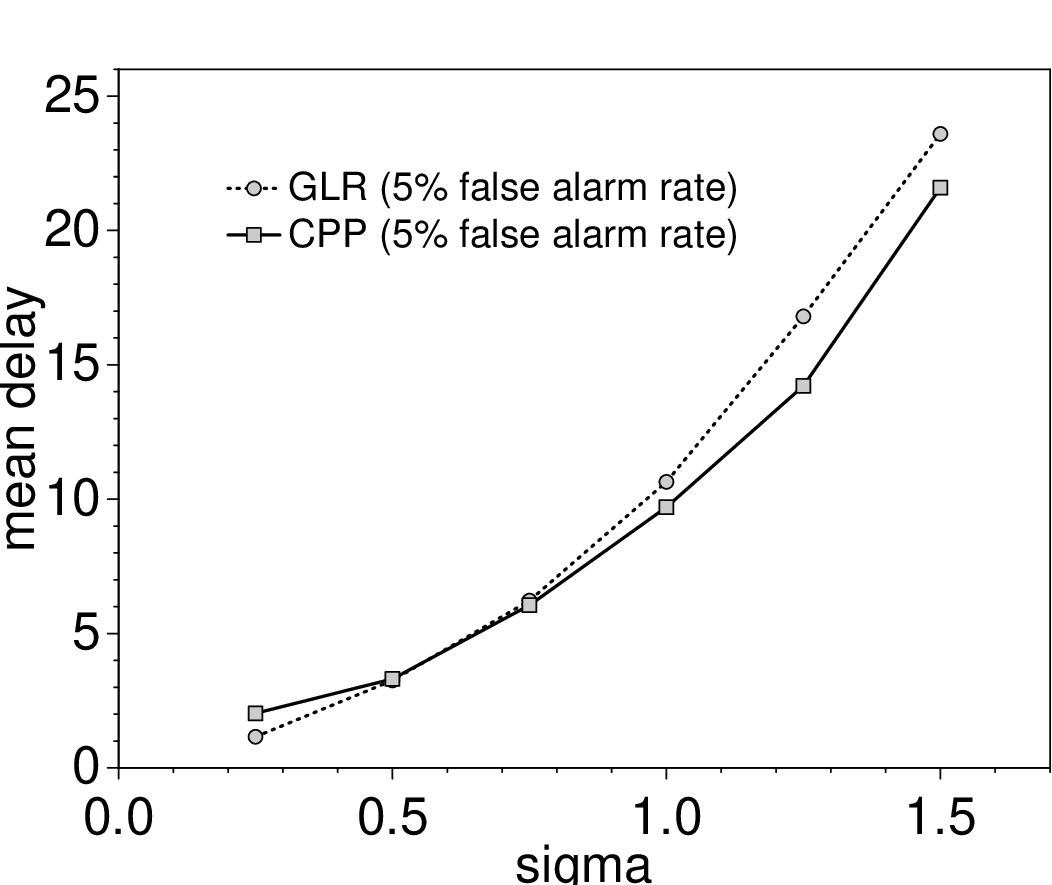}
\caption{Mean delay versus $\sigma$ with false alarm probability
$\alpha = 0.05$.}
\label{matsig}
\end{figure}

The performance of both algorithms depends on the choice of the
threshold.  To compare algorithms, we plot mean delay versus
false alarm probability for a range of thresholds.  Then for a fixed
false alarm probability, $\alpha$, we can see which algorithm
yields a shorter mean delay.

Figure~\ref{matfar} shows the results averaged over 1000 runs.
Both algorithms use the same random number generator, so they
see the same data. 

Across a wide range of $\alpha$, the mean alarm time for CPP is
slightly better than for GLR; the difference is 0.5--1.0 time steps.
For example, at $\alpha = 0.05$, the mean alarm time is 9.7 for CPP and
10.7 for GLR (estimated by linear interpolation between data points).

This advantage increases as $\sigma$ increases (or equivalently as
$S/N$ decreases).  Figure~\ref{matsig} shows the mean alarm time with
$\alpha = 0.05$ for a range of values of $\sigma$.  When $\sigma$ is
small, the mean alarm time is small, and GLR has an advantage.  But
for $\sigma > 0.7$ ($S/N < 1.4$) CPP is consistently better.

We conclude that CPP is as good as GLR, or better, over a wide range
of scenarios, and that the advantage increases as the signal-to-noise
ratio decreases.

\section{Conclusions}

We have proposed a novel algorithm for estimating the location of
the last changepoint in a time series that includes abrupt changes.
The algorithm is Bayesian in the sense that the result is a joint
posterior distribution of the parameters of the process, as opposed
to a hypothesis test or an estimated value.  Like other Bayesian
methods, it requires subjective prior probabilities, but if we
assume that changepoints are equally likely to occur at any time,
the required prior is a single parameter, $f$, the probability
of a changepoint.

This algorithm can be used for changepoint detection, location
estimation, or tracking.  In this paper, we apply it to the
changepoint detection problem and compare it with a standard algorithm
for this problem, GLR.  We find that the performance is as good as
GLR, or better, over a wide range of scenarios.

In this paper, we focus on problems where values between changepoints
are known to be normally distributed.  In some cases, data from other
distributions can be transformed to normal.  For example, several
Internet characteristics are lognormal
\cite{paxson94empirically}, which means that they are normal under a
log transform.  Alternatively, it is easy to extend our algorithm to
handle any distribution whose pdf can be computed, but there is
likely to be a performance penalty (the normal assumption is
convenient in our implementation because its sufficient statistics can
be updated incrementally with only a few operations).

We think that the proposed algorithm is promising, and makes a novel
contribution to the statistics of changepoint detection.
But this work is at an early stage; there is still a lot to do:

\begin{enumerate}

\item We have evaluated the proposed algorithm in the context of the
changepoint detection problem, and demonstrated its use for estimating
the location of a changepoint, but we have not evaluated its
performance for location estimation and tracking.  These problems take
advantage of the full capability of the algorithm, so they will test
it more rigorously.

\item In Section~\ref{varchange} we identified an intermittent problem
when we use our algorithm to detect changes in variance.  We are
working on mitigating this problem.

\item A drawback of the proposed algorithm is that it requires
time proportional to $n^2$ at each time step, where $n$ is the number
of steps since the second-to-last changepoint.  If the time between
changepoints is more than a few thousand steps, it may be necessary
to desample the data to control run time.

\item So far we have not addressed the autocorrelation structure that
has been seen in many network measurements, but this work raises an
intriguing possibility.  If a time series actually contains abrupt
changepoints, but no correlation structure between them, it
will appear to have autocorrelation if we ignore the changepoints.
In future work we plan to use changepoint detection to divide a time
series of measurements into stationary intervals, and then investigate
the autocorrelation structure {\em within} intervals to see whether
it is possible that observed correlations are (at least in part) an
artifact of stationary models.

\end{enumerate}

\bibliographystyle{plain}
\bibliography{paper}

\end{document}